\documentclass[conference]{IEEEtran}
\usepackage[utf8]{inputenc}
\usepackage{enumitem}
\usepackage{hyperref}
\usepackage{cite}

\title{Der Weg zur digitalen Arbeitsmappe\\[1mm] \large Digitales Prüfungswesen mit Zertifizierung\\[2mm]10.07.2024}

\author{
    \IEEEauthorblockN{Martin Becke\IEEEauthorrefmark{1}\textsuperscript{\textdagger}, Julia Padberg\IEEEauthorrefmark{1}\textsuperscript{\textasteriskcentered}}\\
    \IEEEauthorblockA{\IEEEauthorrefmark{1} NahLab! HAW Hamburg\\
    \textsuperscript{\textdagger}Martin.Becke@haw-hamburg.de, \textsuperscript{\textasteriskcentered}Julia.Padberg@haw-hamburg.de}
}

\begin{document}

\maketitle

\begin{abstract}
Ziel der Arbeit ist es, einen alternativen Ansatz zur Erfassung und Bewertung von Studierendenleistung vorzustellen, der eine nachhaltige Leistungserfassung mit der Möglichkeit der Integration von insbesondere Praxisanteilen ermöglicht. Das angestrebte Ergebnis ist eine digitale Mappe mit Arbeitsproben -- und nicht nur Zeugnissen, die im Rahmen der akademischen Bewertung als Portfolioprüfung verstanden werden kann. Dabei geht es mehr um die Erfassung, Bewertung und Zertifizierung von Lernfortschritten und Kompetenzen als um die punktuelle Bewertung einer Leistungsüberprüfung, wie dies  z.B. heute durch die Abgabe von Abschlussarbeiten erfolgt. 
Die Idee ist insbesondere in höheren Semestern Abschlussarbeiten und Leistungstests zu erweitern und später zu ersetzen und stattdessen elektronisch protokollierte Portfolio-Prüfungen - ausgeführt am Beispiel von Lehrprojekten - einzuführen.
\\
Technologisch basiert der Ansatz auf Blockchain und Wallets/Repositories und im weitesten Sinne auf einer Implementierung von Smart Contracts. Dieser Ansatz ist nicht neu, wurde aber bisher noch nicht konsequent für Arbeitsproben in der Informatik durchdacht. Der technologische Ansatz von Smart Contracts ermöglicht ein hohes Maß an Nachvollziehbarkeit und Transparenz bei geringem Verwaltungsaufwand. Darüber hinaus bietet er eine sichere Zertifizierung der Leistungen durch den Anbieter. Klar zu benennen ist, dass weder die Portfolioprüfung noch die Verwaltung von Studienleistungen mit Smart Contracts die originäre Idee ist, sondern die Veränderung der Erfassung von Studienleistungen, hin zu nachvollziehbaren und teilbaren Arbeitsproben. Um den Ansatz über die Arbeitsproben zu motivieren, soll in einem ersten Schritt die aktuelle Problematik mit Abschlussarbeiten diskutiert werden, die auch auf andere Leistungserfassungen übertragbar ist. 
\\
Primäres Ziel dieser Ideenskizze ist die Entwicklung eines individualisierten Leistungsnachweises für Studierende, der aber auch dazu beitragen kann, Leistungen transparenter und nachvollziehbarer zu machen.
\end{abstract}

\section{Die Geschichte der Abschlussarbeiten}

Ursprünglich, um das 12 Jahrhundert, wurden Dissertationen nicht primär vom Kandidaten, sondern von Professoren selbst erstellt, da die wissenschaftliche Forschung und das Verfassen von akademischen Schriften als eine der Hauptaufgaben von Professoren angesehen wurden. Die Dissertation, damals "dissertatio" genannt~\cite{Blair}, diente lediglich als ergänzendes Thesenpapier zur Vorbereitung der Disputation. Dennoch, diese Arbeiten stellten nicht selten Höhepunkte ihrer intellektuellen und wissenschaftlichen Bemühungen dar und waren entscheidend für die Weiterentwicklung ihres Fachgebiets. Studierenden, insbesondere auf niedrigeren Bildungsstufen, wurde die Erstellung solcher tiefgehender Forschungsarbeiten nicht zugetraut, da sie als nicht ausreichend qualifiziert galten, um eigenständig originäre Beiträge zur Wissenschaft zu leisten.~\cite{Schmidt-Kunsemuller1979-vd}

Die Bedeutung der schriftlichen Abschlussarbeit, wie wir sie heute kennen, begann sich im Mittelalter zu entwickeln, insbesondere an den Universitäten in Europa. Ab dem 12. Jahrhundert gewannen Universitäten wie die Universität Bologna und die Universität Paris an Bedeutung. Zu dieser Zeit wurden schriftliche Arbeiten, besonders Dissertationen, ein fester Bestandteil des akademischen Curriculums für fortgeschrittene Studierende, insbesondere für diejenigen, die einen Doktortitel anstrebten. Diese Tradition setzte sich fort und festigte sich im Laufe der Jahrhunderte, wobei die Dissertation heute eine unverzichtbare Voraussetzung für den Erwerb zum Beispiel eines Doktorgrades ist.~\cite{hps_medieval_universities}

Die schriftliche Abschlussarbeit hat sich auch für Master- und Bachelorstudiengänge zu einem zentralen Bestandteil des akademischen Curriculums entwickelt. Ursprünglich waren umfangreiche wissenschaftliche Arbeiten hauptsächlich den Doktoranden vorbehalten, aber im Laufe des 20. Jahrhunderts wurde zunehmend erkannt, dass auch Studierende auf Bachelor- und Masterniveau durch die Erstellung einer Abschlussarbeit wesentliche wissenschaftliche Kompetenzen erwerben können. Auch die Hausarbeiten entwickelten sich erst in diesem Zeitraum zu einer Standardkomponente des Hochschulsystems.

Für Bachelorstudiengänge wurde die Abschlussarbeit im Zuge der Bologna-Reform~\cite{terry2008bologna} und der Einführung gestufter Studiengänge in Europa ab den späten 1990er Jahren zum Standard. Eine Motivation war, einen wissenschaftlichen Kern für die Leistungserfassung zu definieren, der ein hohes Maß an individueller Entwicklung berücksichtigt. Ob diese Motive unter den heutigen Bedingungen noch zutreffen, ist sicherlich eine neue Betrachtung wert. 

\section{Probleme bei Abschlussarbeiten}

Die Konzentration auf die Bewertung der Abschlussarbeiten, insbesondere Bachelor- und Masterarbeiten, weisen eine Reihe von inhärenten Problemen auf, die trotz ihres pädagogischen und wissenschaftlichen Werts häufig kritisiert werden:

\begin{itemize}[leftmargin=*]
    \item \textbf{Betreuungsqualität und -intensität:}
    \newline Die Qualität und Intensität der Betreuung durch Professoren und Dozenten ist sehr unterschiedlich. Einige Studierende werden intensiv betreut, während andere weitgehend auf sich allein gestellt sind. Eine unzureichende Betreuung kann sich stark negativ auf die Qualität der Abschlussarbeit auswirken, da Studierende ohne ausreichende Rückmeldung und Anleitung oft Schwierigkeiten haben, wissenschaftlichen Standards von Anfang an gerecht zu werden. \cite{betreuungsleistung2023},\cite{ghadirian2014}.  Im aktuellen System ist der Prozess im Zeitverlauf für Dritte nicht nachvollziehbar, ebenso wenig wie der Einfluss der Betreuung. 

    \item \textbf{Zeit- und Ressourcenmangel:}
    \newline Studierende haben oft begrenzte Zeit und Ressourcen, um ihre Abschlussarbeit zu verfassen, was zu suboptimalen Ergebnissen führen kann. Insbesondere der Druck, innerhalb eines vorgegebenen Zeitrahmens zu arbeiten, führt häufig zu oberflächlichen oder unvollständigen Analysen. Zudem fehlt den Studierenden manchmal der Zugang zu wichtigen Ressourcen wie Datenbanken und wissenschaftlicher Literatur \cite{Rabab}. Dies ist mitunter sehr abhängig von der Einrichtung. Aber auch mit entsprechender Unterstützung wird teilweise auf Suchmethoden wie die klassische Google-Suche zurückgegriffen, die zwar einen schnelleren, aber nicht immer den qualitativ besten Zugang ermöglichen. Dies gilt in eingeschränkter Form auch für optimierte Suchmaschinen wie scholar.google.com. Quellenarbeit wird immer mehr zu einer oberflächlichen Anwendung von Suchalgorithmen. Die Vergleichbarkeit über die gemeinsame Bearbeitungszeit herzustellen, hat sich als nicht zielführend erwiesen, da dies bereits bei unterschiedlicher Interpretation der Quellenarbeit zu völlig unterschiedlichen Ansätzen führt. Für Dritte ist es wichtiger, nachvollziehen zu können, welche Schritte tatsächlich zum Aufbau der Wissensbasis erbracht wurde. 

    \item \textbf{Uneinheitliche Bewertungskriterien:}
    \newline Die Bewertung von Abschlussarbeiten kann sehr unterschiedlich ausfallen, da es keine einheitlichen Kriterien gibt. Man kann sich sicher auch der Diskussion stellen, ob es diese überhaupt geben sollte. Es führt aber in jedem Fall zu Inkonsistenzen und Subjektivität bei der Notenvergabe. Diese Bewertung ist jedoch notwendig, da sie in ein standardisiertes Bewertungssystem überführt werden muss. Auch wenn man daran festhalten möchte, wäre es vielleicht besser, für die Interessenten - z.B. Arbeitgeber - den Prozess selbst zu dokumentieren und die Bewertung der Leistung dem Interessenten selbst zu überlassen oder anderen Werkzeugen für eine Zusammenfassung bereitzustellen. 

    \item \textbf{Plagiarismus und wissenschaftliche Integrität:}
    \newline Der Druck, eine qualitativ hochwertige Arbeit abzuliefern, kann Studierende dazu verleiten, unethische Mittel wie dem Plagiat zu nutzen. Fälle von Plagiarismus in akademischen Arbeiten, einschließlich Bachelor- und Masterarbeiten, sind bekannt und weiter erwartbar. Dies untergräbt die wissenschaftliche Integrität und den Bildungswert der Arbeiten. \cite{Larsson2013}. Wenngleich dieses Problem nicht unmittelbar gelöst werden kann, können, wenn die Arbeiten in digitaler Form vorliegen, diese Herausforderungen auch in Zukunft jederzeit mit neuen Werkzeugen wieder automatisiert aufgearbeitet werden. Dies kann technisch automatisiert gelöst werden, in dem Repositories der Studierenden wiederkehrend in Prozesse eingepracht werden, die Leistungen mit neuen Werkzeugen untersuchen.

    \item \textbf{Ungleichmäßige Vorbildung und Fähigkeiten:}
    \newline Studierende kommen mit unterschiedlichen Vorkenntnissen und Fähigkeiten in die Abschlussphase ihres Studiums, was sich auf die Vergleichbarkeit und Qualität der Abschlussarbeiten auswirkt. Natürlich ist Exzellenz ein absoluter Maßstab, der auch angestrebt und bewertet werden muss. Hochschulen bereiten jedoch auch auf das Berufsleben vor, wobei andere Aspekte eine wichtige Rolle spielen können, wie z. B. die Etablierung einer Lernkurve.
    Beispielhaft verfassen Studierende mit besserem Zugang zu Vorkenntnissen und zusätzlichen Lernressourcen tendenziell bessere Abschlussarbeiten, was zu strukturellen Ungleichheiten führt \cite{Evaluating2022}. Hier sollte beispielhaft noch mehr auf den absoluten und den relativen Lernerfolg eingegangen werden. 
    Der absolute Lernerfolg misst die Gesamtmenge an Wissen und Fähigkeiten, die ein Studierender während des Studiums erworben hat. Dieser ist jedoch stark vom Vorwissen abhängig, welches oft nicht bekannt oder protokolliert ist. Im Gegensatz dazu bezieht sich der relative Lernerfolg auf die Fortschritte, die ein Studierender im Verhältnis zu seiner Ausgangsposition innerhalb eines bestimmten Zeitraums macht. Dieser relative Fortschritt kann ein aussagekräftigeres Maß für den individuellen Lernerfolg sein, da er die unterschiedlichen Ausgangsbedingungen der Studierenden berücksichtigt. Die Fokussierung auf den relativen Lernerfolg kann auch dazu beitragen, die Auswirkungen struktureller Ungleichheiten besser zu verstehen und gezieltere Maßnahmen zur Unterstützung der Studierenden zu entwickeln, wodurch auch die absolute Exzellenz besser erreicht werden kann.

    \item \textbf{Innovation und Originalität:}
    \newline Die Konzentration auf formale Kriterien können Studierende davon abhalten, risikoreiche und innovative Forschung zu betreiben. Die Möglichkeit, auch ein Scheitern auf hohem Niveau zu dokumentieren, kann einen erheblichen Mehrwert darstellen, der Innovation und Originalität fördern kann.
\end{itemize}

Die Motivation dieser Arbeit liegt nicht im Versuch der Behebung der Unzulänglichkeiten, sondern vielmehr in der Ermöglichung einer umfassenderen und systematischeren Leistungserfassung für Bachelor- und Masterleistungen um die strukturellen Nachteile auszugleichen. Anstatt die inhärenten Nachteile direkt zu beheben, soll mehr Transparenz durch mehr Möglichkeiten zur detaillierten Protokollierung von Arbeits- und Lernprozessen geschaffen werden. Dies kann nicht nur dem Feedback der Dozenten genügen, sondern auch dem der anderen Studierenden. Dabei steht der Prozess selbst im Vordergrund und nicht die Abbildung auf eine standardisierte Leistungsbewertung, die schwer umsetzbar ist und nicht existiert. Im besten Fall wird sogar eine individuelle Benotung überflüssig, da sich jeder Interessierte selbst ein Bild von den Leistungen machen kann bzw. durch technische Hilfsmittel nach eigenen Präferenzen erstellen lassen kann. Die Werkzeuge für eine bedarfsgerechte Leistungsbewertung könnten allgemein auch mit KI umschrieben werden.

\section{Neue Herausforderung durch KI}

Eine neue Herausforderung im Kontext der Abschlussarbeit liegt in der Einführung immer umfassender KIs, aktuell insbesondere Large Language Models (LLMs). Die aktuellen Auswirkungen von KI-Werkzeugen wie OpenAI ChatGPT, Google Gemini und Perplexity auf die Erstellung von Abschlussarbeiten sind vielfältig und  noch nicht vollständig untersucht. Sie betreffen aber verschiedene Aspekte des akademischen Prozesses und sie werden bei der Erstellung von Abschlussarbeiten eingesetzt, ob verboten oder nicht. Sie können hilfreiche Werkzeuge sein, auch für den Prozess der Entwicklung einer Abschlussarbeit, dazu gehören:

\begin{itemize}[leftmargin=*]
    \item \textbf{Erleichterter Zugang zu Informationen:}
    \newline KI-Werkzeuge ermöglichen den Studierenden einen schnellen Zugriff auf umfangreiche Datenbanken, wissenschaftliche Artikel und Fachinformationen, was die Recherchezeit zur Informationsbeschaffung erheblich verkürzen kann, im Zweifelsfall aber auf die trainierte Datenmenge beschränkt bleibt.
    \newline Richtig angewandt, können diese Tools  aktuelle Informationen liefern, die für zeitkritische Themen relevant sind.

    \item \textbf{Verbesserte Schreib- und Bearbeitungsprozesse:}
    \newline Textgenerierung und -verbesserung: ChatGPT, DeepL write und ähnliche Modelle können beim Formulieren von Texten, Strukturieren von Argumenten und Korrigieren von sprachlichen Fehlern helfen, um  die Qualität der Arbeiten zu steigern.
    \newline Insbsondere kann  KI  Studierende unterstützen, die in einer Fremdsprache schreiben, indem sie sprachliche Korrekturen und stilistische Verbesserungen vorschlägt.

    \item \textbf{Lernunterstützung und Verständnisförderung:}
    \newline KI-Tools bieten individuelle Erklärungen und Lernhilfen, die das Verständnis komplexer Themen verbessern.
    \newline Diese Werkzeuge können als virtuelle Tutoren fungieren und detaillierte Anleitungen zu spezifischen Fragestellungen liefern.
\end{itemize}

Eine KI ist in diesem Zusammenhang aus Sicht der Autoren ein unterstützendes Werkzeug, vergleichbar mit einer Rechtschreibprüfung, einem Taschenrechner oder einer Graphensuche in einer Wissensdatenbank. Diese Werkzeuge können für eine bestimmte Leistung ausgeschlossen werden, insbesondere in der Grundausbildung, aber ein Ausschluss für den Berufs- und Forschungsalltag ist unrealistisch und auch nicht zielführend. Inwieweit hier die Nutzung transparenter gestaltet werden muss als bei den bisherigen Werkzeugen, kann auch mit dem hier vorgeschlagenen Ansatz diskutiert werden, stellt aber nicht den Problemraum und die Motivation für diese Ideenskizze dar.

\section{Herausforderungen für die Bewertung}

Im Zuge der Diskussion um KI als Werkzeug müssen aber auch die Herausforderungen dieser Werkzeuge für den Bewertungsprozess diskutiert werden.

\begin{itemize}[leftmargin=*]
    \item \textbf{Weiter erhöhte Gefahr des Plagiats:}
    \newline Unethische Nutzung: Studierende könnten KI-generierte Texte ohne angemessene Zitierung als ihre eigenen ausgeben, wodurch die akademische Integrität der Arbeit gefährdet wäre.
    \newline Auch wird es schwieriger, die Originalität der Arbeiten zu überprüfen, da KI-generierte Inhalte oft gut formuliert und schwer zu identifizieren sind. Je loser das Betreuungsverhältnis ist, desto schwieriger ist die Beurteilung. 

    \item \textbf{Übermäßige Abhängigkeit und Mangel an Eigenständigkeit:}
    \newline Eine zu starke Abhängigkeit von KI-Werkzeugen kann dazu führen, dass Studierende weniger eigenständig arbeiten und ihre kritischen Denkfähigkeiten, insbesondere in der Grundlagenausbildung, nicht ausreichend entwickeln.
    \newline Studierende könnten daher die grundlegenden Fähigkeiten zur Durchführung eigenständiger Forschung vernachlässigen. Es sollte daher nachvollziehbar sein, für welche Leistungen, welche Werkzeuge erlaubt waren.

    \item \textbf{Qualität und Verlässlichkeit der Informationen:}
    \newline Ungenaue oder veraltete Daten: KI-Modelle können fehlerhafte oder ungenaue Informationen liefern, was die Qualität der wissenschaftlichen Arbeiten beeinträchtigen kann.
    \newline Bias und Fehlinterpretationen sind zu erwarten. Künstliche Intelligenz kann Verzerrungen enthalten oder Informationen falsch interpretieren, was zu falschen Schlussfolgerungen führen kann.  Dies gilt zwar auch für die klassische Google-Suche, allerdings werden hier alte Ergebnisse mit neuen vermischt, in einem KI-veralteten Modell werden nur die veralteten Daten aufbereitet. 

    \item \textbf{Ethische und rechtliche Bedenken:}
    \newline Datenschutz und Sicherheit sind auch relevante Themen. Der Einsatz von KI in der akademischen Arbeit wirft Fragen zum Datenschutz und zur Sicherheit der verwendeten Daten auf.
    \newline Die Nutzung von KI-generierten Inhalten kann urheberrechtliche Fragen aufwerfen, insbesondere wenn die Quellen der Informationen nicht klar sind, was auch für eine Veröffentlichung über die Hochschulen Fragen aufwirft.
\end{itemize}

Viele akademische Einrichtungen überarbeiten derzeit ihre \cite{mrass2023chatgpt} Richtlinien, um den Einsatz von KI in Abschlussarbeiten zu regeln und die Einhaltung ethischer Standards sicherzustellen. Die hier formulierte These ist, dass diese Anforderungen mit den derzeitigen Bewertungssystemen nicht gut abgebildet werden können. Weiter bieten Hochschulen vermehrt Schulungen an, um Studierende und Lehrende über den verantwortungsvollen Umgang mit KI-Tools zu informieren. Somit kann man sicher feststellen, dass verstärkt daran gearbeitet wird, KI-Werkzeuge in bestehende Lernmanagementsysteme zu integrieren, um den Studierenden eine nahtlose Unterstützung zu bieten. Die Herausforderung ist nur diese bieten keinen langfristigen Lösungsansatz, insbesondere nicht für die Festlegung einer nachvollziehbaren Bewertungsgrundlage. Hier müssen mehr Informationen über die Entstehung der Prüfungsleistung gehalten werden.
Diese Anforderung ist eng mit der technischen Umsetzung der Erfassung der Ergebnisse der Arbeitsschritte verbunden, weshalb diese Arbeit auch einen Einblick in die technische Umsetzung geben soll um das Gesamtbild zu beleuchten. 

\section{Ansatz}

Die Ergänzung oder gar Ablösung des Notensystems in den höheren Semestern\footnote{Eine Art Grundstudium über 2-3 Semester sollte dem System vorgeschaltet sein.} durch ein Portfolio-Prüfungssystem mit starkem Anwendungsbezug und der Speicherung der Ergebnisse in einem elektronischen Wallet, das durch Smart Contracts auf einer öffentlichen Blockchain zertifiziert wird, bietet eine zeitgemäße und praxisorientierte Alternative, von der Studierende, Forscher und Arbeitgeber gleichermaßen profitieren. Es muss aber auch noch einmal betont werden, dass nach Überzeugung der Autoren die Grundausbildung von diesem Ansatz ausgenommen werden muss und der Ansatz nur die höheren Semester betrifft. 

Die Vorteile des Portfolio-Prüfungssystems, insbesondere auch für eine Hochschule für angewandte Wissenschaften (HAW), können im anwendungsorientierten Lernen gesehen werden. Grundsätzlich ist bekannt, dass Portfolio-Prüfungen Studierenden ermöglichen, ihre Fähigkeiten durch reale Projekte und Anwendungen zu demonstrieren, die direkt auf ihre zukünftigen beruflichen oder wissenschaftlichen Tätigkeiten zugeschnitten sind. Der besondere Ansatz der kontinuierlichen elektronischen Nachverfolgung erlaubt weiter eine transparente Bewertung. Anstatt punktuelle Bewertungen durch Klausuren und Abschlussarbeiten, bietet ein kontinuierliche Portfolio von praxisbezogenen Leistungen einen umfassenden Überblick über die Leistungen der Studierenden.

Im besten Fall wird nicht die Optimierung auf eine Prüfungsform gesucht, sondern die Optimierung am Projekt mit den einhergehenden Schlüsselkompetenzen. Studierende werden dazu angeregt, komplexe Probleme eigenständig zu analysieren und zu lösen, was ihre kritischen Denkfähigkeiten stärkt, und das Bewertungssystem auch den Nachweis von verschiedenen Kompetenzen am Projekt zulässt. In einem Informatik-Projekt könnte dies exemplarisch sein:

\begin{itemize}[leftmargin=*]
    \item \textbf{Projektergebnisse:}
    \newline Softwareentwicklung: Entwicklung einer Softwarelösung von der Konzeption bis zur Implementierung.
    \newline Teamprojekte: Teilnahme an kollaborativen Projekten, die Teamarbeit, Kommunikation und Projektmanagementfähigkeiten fördern.

    \item \textbf{Praktische Übungen:}
    \newline Programmierung: Lösung von Programmieraufgaben und -problemen in verschiedenen Programmiersprachen.
    \newline Laborübungen: Durchführung und Dokumentation von praktischen Übungen im Labor, z.B. Netzwerkkonfiguration, Datenbankverwaltung oder Hardware-Tests.

    \item \textbf{Forschungsarbeiten:}
    \newline Literaturrecherche: Erstellung eines wissenschaftlichen Berichts zu einem aktuellen Thema der Informatik, basierend auf einer umfassenden Literaturrecherche.
    \newline Empirische Studien: Durchführung und Auswertung von Studien, Experimenten oder Umfragen zu informatikbezogenen Fragestellungen.

    \item \textbf{Präsentationen (mit Aufzeichnung):}
    \newline Mündliche Präsentationen: Vorstellung von Projekten, Forschungsergebnissen oder spezifischen Themen vor einer Gruppe oder einem Prüfungskomitee.
    \newline Poster-Sessions: Erstellung und Präsentation von wissenschaftlichen Postern zu Forschungsprojekten oder praktischen Arbeiten.

    \item \textbf{Dokumentationen und Berichte:}
    \newline Technische Dokumentation: Ausführliche Dokumentation von Softwareprojekten, inklusive Quellcode, Architektur, Designentscheidungen und Benutzerhandbücher.
    \newline Projektberichte: Detaillierte Berichte über den Verlauf und die Ergebnisse von Projekten, inklusive Reflexion über die eigene Arbeit und die Teamdynamik.

    \item \textbf{Kleine Tests/Vorprüfungen:}
    \newline Zwischenprüfungen: Kurze schriftliche oder mündliche Prüfungen zu bestimmten Themenbereichen, um das Wissen und Verständnis zu überprüfen.
    \newline Online-Quizzes: Regelmäßige, formative Tests, die den Lernfortschritt der Studierenden kontinuierlich überprüfen.

    \item \textbf{Reflexionsberichte:}
    \newline Selbstreflexion: Berichte, in denen Studierende ihre Lernprozesse, Fortschritte und Herausforderungen reflektieren und evaluieren.
    \newline Feedback-Integration: Analyse und Integration des erhaltenen Feedbacks von Dozenten und Mitstudierenden.

    \item \textbf{Peer-Reviews:}
    \newline Bewertung von Mitstudierenden: Teilnahme an Peer-Review-Prozessen, bei denen Studierende die Arbeiten ihrer Kommilitonen bewerten und konstruktives Feedback geben.

    \item \textbf{Zusatzqualifikationen:}
    \newline Zertifikate: Erworbenen Zusatzqualifikationen oder Zertifikate in spezifischen Technologiebereichen (z.B. AWS, Google Cloud, Cisco).
    \newline Workshops und Seminare: Teilnahme und aktive Mitwirkung an relevanten Workshops, Seminaren oder Konferenzen.
\end{itemize}

\subsection{Technik ist auch keine Lösung}
An dieser Stelle sei noch einmal darauf hingewiesen, dass die technische Erfassung nicht den Umfang der Leistungsnachweise erhöht, sondern nur den Raum für die Erfassung der Historie, des Vorgehens und der Qualität der praktischen Umsetzung.
Letztlich wird das System die Probleme der individuellen Leistungsbeurteilung nicht lösen, aber es wird ein vollständigeres Bild liefern, als es die klassischen Noten von 0 bis 15 können. Insbesondere die größere Transparenz in der Entstehung der Leistung als bei einer Mappe von Arbeitsproben erhöht den Nutzen und die Aussagekraft. 
Die Technik selbst wird in einem späteren Kapitel nur als Vorschlag behandelt, für die inhaltliche Diskussion kann aber bereits unter \cite{Idee} oder \cite{brazil} ein vergleichbarer technischer Ansatz nachvollzogen werden, um die Umsetzbarkeit zu validieren. Die Technik ist stark verbunden mit der Struktur. Dennoch sollte eine technische Kritik an der Umsetzung  nicht die Vorteile des Konzepts in Frage stellen. Das grundsätzlich diese Art der technischen Umsetzung bereits ihren Einsatz finden, können verschiedene Quellen wie \cite{blockcerts2024} oder \cite{blockchaincert2024} nachweisen. Hier liegt aber eindeutig der Fokus auf der Verwaltung von Zertifikaten, nicht die Dokumentation eines Lernprozesses.  
\\\\
Auch sollte die Leistungserfassung nicht mit der Beschreibung eines Studienganges verwechselt werden. Um beispielhaft am Ende ein ausgewogenes Informatikstudium zu gewährleisten, könnten bestimmte Leistungen als obligatorisch oder mit einer gewünschten Gewichtung in einem Smart Contract beschrieben werden. Die tatsächliche Beschreibung eines Studienganges in einem Smart Contract soll an dieser Stelle einer weiteren Diskussion überlassen werden soll. Auch soll das klassische Notensystem nicht völlig verdrängt werden. Regeln zur Überführung in das klassische Bewertungssystem sind denkbar. Festzuhalten für diese Ideenskizze ist, dass der Schwerpunkt nun auf Methoden und Kompetenzen und nicht primär auf Modulen liegt. Im Zweifelsfall können die Studierenden sogar sehr unterschiedliche Arbeitsproben in ihrem Portfolio sammeln. Auch ob diese für einen akademischen Titel ausreichend sind, kann sogar individuell von einem Gremium erst nach einer Vorlage beurteilt werden, und muss nicht zwingend im Prozess oder im Smart Contract selbst beschrieben sein. 
\subsection{Technik ist ein Teil der Lösung}
Durch den neuen Fokus auf die Arbeitsprobe und nicht auf die schriftliche Abschlussarbeit oder Abschlussprüfung besteht die Chance, dass mehr kreative und innovative Projekte entwickelt werden, die auch in der modernen Arbeitswelt gefragt sind. Die Lehrenden können detailliertes und individuelles Feedback geben, das den Studierenden hilft, ihre Stärken und Schwächen besser zu verstehen und gezielt daran zu arbeiten. Dieses Feedback und die Reaktionen darauf können digital erfasst werden. Die Bewertung erfolgt auf mehreren Ebenen (fachlich, methodisch, sozial, aber auch zeitlich), so dass ein umfassenderes Bild der Kompetenzen der Studierenden entsteht.
\\\\
Ein wichtiger Teil der Kernidee ist die digitale Signierung/Zertifizierung vorhandener Leistungen mit der Bewertung durch die Dozenten.  Technologisch ist diese Idee nicht neu und wird bereits in verschiedenen Kontexten verwendet. Beispiele finden sich in \cite{alammary2019blockchain}. Diese Ansätze konzentrieren sich jedoch hauptsächlich auf die Schaffung eines alternativen Zertifizierungssystems. Diese Ansätze sind gut und können bei einer ersten Evaluierung für die Ideenskizze sehr hilfreich sein. Technologisch ist  es durch die Integration von Blockchain-Technologien mit Ansätzen aus dem Kontext von Smart Contracts beschrieben. Durch die öffentliche Persistierung der Dozentenzertifikate und der damit nachgewiesenen erbrachten Leistungen in einer öffentlichen Blockchain wird die Integrität und Unveränderbarkeit der Daten sichergestellt. So werden Manipulationen verhindert und auch eine langfristige Historie ermöglicht. Die Leistungen selbst können die Studierenden in einer elektronischen Wallet/Repository speichern, welches ihre Persönlichkeitsrechte schützt. In dem hier vorgeschlagenen Modell können Studierende jedoch ausgewählten Dritten Zugang zu ausgewählten Informationen gewähren.  So können beispielsweise Arbeitgeber anhand der öffentlich zugänglichen Zertifikate schnell und zuverlässig die Echtheit der vorgelegten Arbeiten überprüfen, und so  den Bewerbungsprozess effizienter und nachvollziehbarer gestalten. Erste Ideen dieses Ansatzes sind beispielhaft in \cite{Idee} besprochen und werden im Zuge dieser Ideenskizze noch technisch detaillierter diskutiert.
\\\\
Die Blockchain-Technologie schafft Vertrauen bei zukünftigen Arbeitgebern, da die Zertifikate fälschungssicher sind und die Leistungen der Bewerber verlässlich und viel konkreter widerspiegeln. Langfristig ist es auch von großem Nutzen, wenn die Kompetenzen in digitaler Form vorliegen, um die Inhalte in Zukunft automatisiert gegen überladene Plagiate härten zu können, auch wenn diese Werkzeuge heute noch nicht zur Verfügung stehen. Dies erhöht potenziell die Hemmschwelle, das derzeit schwache System auszunutzen. Auch die Möglichkeiten automatisierter Optimierungswerkzeuge erscheinen Endlos und gehen über Suche, Lernerfolgskontrolle bis hin zu Schwachstellenanalyse. 

Eine Veränderung des Bewertungssystems in den höheren Semestern hin zu elektronisch erfassten Portfolio-Prüfungen mit Blockchain-Zertifizierungen stellt natürlich einen neuen, innovativen Ansatz dar, der neue Wege geht und für Akzeptanz werben muss. Dennoch fördern die Anwendungsorientierung und kontinuierliche Bewertung wesentliche Kompetenzen und ein vollständigeres Bild der Leistungen, während die Integration von Blockchain-Technologie Sicherheit, Transparenz und Vertrauen in die erbrachten Leistungen schafft.

\section{Technologischer Ansatz}
Die Implementierung eines Portfolio-Prüfungssystems mit Blockchain-Zertifizierung und elektronischen Wallets erfordert eine Integration mehrerer moderner Technologien. Dieser Ansatz soll zunächst auf Umsetzbarkeit im NAhLab! evaluiert werden.\\ Die erste Anwendung kann das neue Konzept in der Leistungserbringung im NahLab! sein.
Die Leistungen der Studierenden im Nahlab! werden dadurch stärker gewürdigt und stellen einen Nachweis ihrer Qualifikation im Bereich der Nachhaltigkeitsthemen der Informatik dar. Studierende des NahLab! ziehen Motivation daraus, dasss ihre Anstrengungen anerkannt werden.
Darüber hinaus werden sie so auch ggf.~für künftige Arbeitgeber interessanter.

Um dieses technische System für die Ideenskizze weiter verständlicher zu machen, folgt nun eine Vorstellung des  technischen Lösungsraums.

\subsection{Elektronische Portfolio-Plattform}
Zunächst kann man die erste Idee der Portfolio-Plattform in einer klassischen Hexagonal Architektur \cite{tanenbaum2017distributed} verstehen. Noch mehr abstrahiert, kann man das Gesamtsystem auch in drei Ebenen beschreiben, bestehend aus einem \textbf{Frontend}, der \textbf{Anwendungslogik} und die \textbf{Persistierung}. Flankiert wird dieses System von einer öffentlichen Blockchain, die beispielhaft in der Domain der HAW Hamburg initiiert wird. 

\subsubsection{Frontend}
Das Frontend besteht in der ersten Idee aus einer \textbf{Wallet-App}, einer \textbf{Wallet-Zertifikats-App} und einer benutzerfreundlichen \textbf{Webanwendung}, die den Studierenden ermöglicht, ihre Arbeiten hochzuladen, zu organisieren und zu präsentieren. 

Die \textbf{Wallet-App:} ist die technische Repräsentation der Arbeitmappe. Die Grundlegende Idee basiert auf Arbeiten wie \cite{Idee}, die ein Blockchain-basiertes Educational Record Repository vorschlagen. Auch in dieser Arbeit sollen ähnliche Eigenschaften wie ein privates Git Repository eingebracht werden, dazu gehören insbesondere folgende Eigenschaften: 
\begin{itemize}
     \item Zugriffsbeschränkung
    \begin{itemize}
     \item \textbf{Private}: Der Zugriff ist auf bestimmte Benutzer beschränkt, die vom Eigentümer oder Administrator (PAV) eingeladen werden.
     \item \textbf{Benutzerrollen}: Unterschiedliche Rollen (z.B. Admin, Maintainer, Contributor) können mit spezifischen Berechtigungen versehen werden.
    \end{itemize}

    \item Authentifizierung und Autorisierung
    \begin{itemize}
     \item \textbf{SSH-Schlüssel}: Verwendung von SSH-Schlüsseln zur sicheren Authentifizierung.
     \item \textbf{OAuth und SSO}: Integration von OAuth und Single Sign-On (SSO) für zentrale und sichere Anmeldung.
     \item \textbf{Zwei-Faktor-Authentifizierung (2FA)}: Zusätzliche Sicherheitsebene durch 2FA.
    \end{itemize}

    \item Audit-Logs
    \begin{itemize}
     \item \textbf{Protokollierung}: Detaillierte Protokolle aller Aktivitäten im Repository.
     \item \textbf{Compliance}: Unterstützung für Compliance-Anforderungen 
    \end{itemize}

    \item Backup und Wiederherstellung
    \begin{itemize}
     \item \textbf{Automatische Backups}: Regelmäßige Backups des Repositorys zur Sicherstellung der Datenintegrität.
     \item \textbf{Wiederherstellung}: Möglichkeit, das Repository im Falle eines Datenverlusts wiederherzustellen.
    \end{itemize}

    \subsection{Speicherverwaltung}
\begin{itemize}
    \item \textbf{Repository-Größe}: Effiziente Verwaltung von großen Repositories durch delta-basierte Speicherung und Kompression.
     \item \textbf{Large File Storage (LFS)}: Unterstützung für die Speicherung und Verwaltung großer Binärdateien.
    \end{itemize}

\end{itemize}

Diese Arbeit basiert auf vergleichbare Bestrebungen am MIT. Das MIT hat ein System zur Erstellung von Blockchain-basierten Anwendungen entwickelt, das offizielle Dokumente ausstellt und verifiziert und als \textit{Blockcerts Wallet} bezeichnet wird. Dieses System ermöglicht beispielsweise die Erstellung eines Zertifikats-Wallets, in dem Studierende ihre virtuellen Diplome über ihre Smart-Geräte empfangen können. Im Gegensatz zum \textit{Blockchain-based Educational Records Repository (BcER2)}, ist das \textit{MIT Blockcerts Wallet} System eine Plattform zur Erstellung von Anwendungen, die eine ähnliche Funktionalität bietet. Aber auch andere Systeme wurde bereits diskutiert, zum Beispiel in \cite{alammary2019blockchain}.
Aus diesen Vorarbeiten lassen sich auch Anforderungen an eine \textbf{Wallet-Zertifikats-App:} ableiten.

Die \textbf{Wallet-Zertifikats-App} dient den Studierenden, aber auch Externen, zur Überprüfung der Zertifikate. Es ist eine (mobile) Web-Applikation, die es ermöglicht Zertifikate und akademischen Leistungen entsprechend vorhandener Profile aufzurufen und zu verifizieren. Die Wallet-Zertifikats-App kann direkt aus der Wallet-App heraus aufgerufen werden. Hier können unterschiedliche Aspekte mit unterschiedlichen Technologien eine Rolle spielen. 

Die Webanwendung muss geeignet sein die Portfolio-Prüfung zu organisieren und zu verwalten. Hierzu gehören eine Reihe an Funktionalitäten, die hier nur angedeutet, aber in dieser Diskussion nicht vollständig diskutiert werden soll. 
\begin{itemize}
    \item \textbf{Dokumentenmanagement:} Hochladen, Bearbeiten und Organisieren von Dokumenten.
    \item \textbf{Versionierung:} Verfolgen und Verwalten verschiedener Versionen von Dokumenten.
    \item \textbf{Feedback und Bewertungen:} Ermöglichen von Kommentaren und Bewertungen durch Dozenten und Peers.
    \item \textbf{Analyse:} Tools zur Analyse der Fortschritte der Studierenden.
\end{itemize}

\subsubsection{Backend}
Eine robuste Server-Infrastruktur zur Verwaltung der Daten, Authentifizierung der Benutzer und Sicherstellung der Datenintegrität.

Zentraler Baustein ist das API Gateway. Das API Gateway kann als zentrale Schnittstelle dienen, um Anfragen von Nutzern entgegenzunehmen und an die entsprechenden Dienste weiterzuleiten. Das System wird auch eingesetzt für die Verwaltung der Benutzeranmeldungen und Authentifizierungen, beispielhaft einschließlich der Zwei-Faktor-Authentifizierung (2FA). Sie bietet auch die Schnittstelle zu der Persistierung der Daten in einer angemessenen Datenbank und fungiert als Transaktionsmonitor.

Jede Zertifizierung oder Prüfungsrelevante Leistung wird als Transaktion auf einer zu bestimmenden Blockchain gespeichert und von einem Smart Contracts verwaltet. Der Smart Contract automatisiert und sichert den gesamten Prozess der Ausstellung, Speicherung und Verifizierung von Zertifikaten, wodurch Transparenz, Sicherheit und Nachverfolgbarkeit gewährleistet werden.

Der Prozess beginnt, wenn ein Studierender eine protokollierte oder begleitete Leistung abgeschlossen hat. Der zuständige Dozent initiiert die Erstellung eines digitalen Zertifikats, das beispielhaft die relevanten Informationen wie den Namen des Studierenden, den Kurs, das Datum und die Bewertung enthält. Der Smart Contract auf der Blockchain übernimmt dann die Ausstellung des Zertifikats. Der Dozent signiert das Zertifikat digital, und der Smart Contract speichert es als unveränderliche Transaktion auf der Blockchain. Diese Speicherung garantiert die Integrität und Unveränderlichkeit des Zertifikats, da jede Änderung einer Blockchain öffentlich nachvollziehbar ist.

Ein wesentlicher Vorteil dieses Systems ist die einfache Verifizierung der Zertifikate durch externe Parteien wie potenzielle Arbeitgeber. Jeder Eintrag auf der Blockchain ist öffentlich zugänglich, und durch einen auf dem Zertifikat enthaltenen QR-Code oder Hash können Arbeitgeber direkt auf die Blockchain-Daten zugreifen, um die Echtheit des Zertifikats zu überprüfen. Dies schafft Vertrauen und reduziert das Risiko von Betrug, da die Daten unveränderlich und transparent sind.

Darüber hinaus bietet die Automatisierung durch den Smart Contract erhebliche Effizienzgewinne. Der Prozess der Ausstellung und Verifizierung von Zertifikaten wird stark vereinfacht und beschleunigt, wodurch administrative Aufgaben reduziert und Fehler minimiert werden. Studierende profitieren von lebenslangem Zugriff auf ihre Zertifikate, die sie bei Bedarf jederzeit präsentieren können, sei es für Bewerbungen oder weitere akademische Zwecke.

\subsubsection{Persistierung der Daten}

Im Bereich der Persistierung sind mehrere Maßnahmen von entscheidender Bedeutung, um die Integrität und Vertraulichkeit der Daten zu gewährleisten. Die Datenverschlüsselung spielt hierbei eine zentrale Rolle. Bei der Übertragung von Daten wird Ende zu Ende Verschlüsselung vorausgesetzt, um eine sichere Datenübertragung zu gewährleisten und Abhörversuche zu verhindern. Ebenso wichtig ist die Verschlüsselung der gespeicherten Daten, die sowohl in den Datenbanken als auch als Kopie in den elektronischen Wallets durchgeführt wird, um die Daten vor unbefugtem Zugriff zu schützen.

Die Einhaltung von Datenschutzrichtlinien stellen sicher, dass die Handhabung von Nutzerdaten den gesetzlichen Vorgaben entspricht. Eine zentrale Anforderung ist die Einhaltung der Datenschutzbestimmungen der EU, bekannt als General Data Protection Regulation (GDPR). Diese Vorschriften stellen sicher, dass personenbezogene Daten der Nutzer sicher und verantwortungsvoll verarbeitet werden. Zudem wird, wo immer möglich, eine Anonymisierung sensibler Daten vorgenommen, um die Privatsphäre der Nutzer weiter zu schützen und die Risiken eines Datenmissbrauchs zu minimieren.

Ein weiteres wichtiges Element ist die Unterstützung von Standards wie Learning Tools Interoperability (LTI), die eine problemlose Integration mit Lernmanagementsystemen (LMS) ermöglichen.

Die Datenmigration umfasst Strategien zur Überführung von Altdaten aus bestehenden Systemen in das neue Portfolio- und Zertifizierungssystem. Dies gewährleistet, dass alle relevanten historischen Daten erhalten bleiben und nahtlos in das neue System integriert werden um auch ein Migration zu ermöglichen. Ein kontinuierlicher Einsatz von Continuous Integration/Continuous Deployment (CI/CD) Pipelines ermöglicht eine automatisierte und wiederholte Prüfung der Datenintegrität und Sicherung. Diese Pipelines unterstützen auch die fortlaufende Verbesserung und Bereitstellung der Plattform, indem sie regelmäßige Updates und automatische Tests ermöglichen, was die Stabilität und Sicherheit des Systems langfristig gewährleistet.

\subsection{Nutzerfreundlichkeit und Zugänglichkeit}

Die Plattform erfüllt nicht nur rein technische Anforderungen, sondern legt auch großen Wert auf Benutzerfreundlichkeit und Unterstützung der Nutzer. Ein benutzerzentriertes Design steht im Mittelpunkt der Entwicklung der Benutzeroberfläche. Diese Oberfläche ist intuitiv gestaltet, um eine einfache Navigation und Nutzung zu ermöglichen. Durch klare, leicht verständliche Layouts und Funktionen wird sichergestellt, dass die Nutzer schnell und effizient mit der Plattform interagieren können.

Ein weiterer wichtiger Aspekt ist die Barrierefreiheit. Die Plattform hält sich an die Web Content Accessibility Guidelines (WCAG), um sicherzustellen, dass sie für alle Nutzer zugänglich ist, einschließlich Personen mit Behinderungen. Dies umfasst Maßnahmen wie die Bereitstellung von Textalternativen für visuelle Inhalte, eine leicht lesbare Textgestaltung und die Kompatibilität mit Hilfstechnologien.

Die technische Umsetzung der Architektur muss aber in ein Gesamtkonzept eingebaut werden. Schulungen und Support sind wichtig. Umfassende Anleitungen und Dokumentationen müssen bereitgestellt werden, damit die Nutzer sich leicht in die Funktionsweise der Plattform aneignen können. Diese Ressourcen erklären die verschiedenen Funktionen und bieten Schritt-für-Schritt-Anleitungen, um typische Aufgaben zu erledigen.

Zur weiteren Unterstützung der Nutzer werden spezielle Support-Teams eingerichtet. Diese Teams stehen bereit, um bei technischen Problemen und Fragen zu helfen. Sie bieten schnellen und effizienten Support, um sicherzustellen, dass die Nutzer die Plattform ohne größere Unterbrechungen nutzen können. Durch diesen ganzheitlichen Ansatz, der sowohl technische Exzellenz als auch Benutzerfreundlichkeit und Unterstützung umfasst, wird die Plattform zu einem leistungsfähigen Werkzeug, das den Bedürfnissen aller Nutzer gerecht wird.

\subsection{Einbindung von KI}
Die Einbindung eines Large Language Models (LLM) wie GPT-4 in das beschriebene System könnte die Benutzererfahrung erheblich verbessern, indem es erweiterte Analysen und kontextbezogene Hilfestellungen bietet.

Das LLM würde als separater Microservice innerhalb der bestehenden Systemarchitektur fungieren. Es könnte über APIs angesprochen werden, um Analysen durchzuführen und kontextbezogene Antworten zu geben.

Das LLM kann komplexe Datenanalysen durchführen, wie z.B. das Erkennen von Trends in den Prüfungsleistungen der Studierenden oder das Identifizieren von häufigen Problemen bei der Verwendung der Plattform. Weiter kann das LLM  Vorhersagen treffen und Empfehlungen basierend auf historischen Daten geben.
Auch kann es auf Anfragen der Nutzer antworten und detaillierte, kontextbezogene Hilfestellungen bieten, insbesondere durch die  Integration von interaktiven Tutorials, die durch das LLM generiert und personalisiert werden können, basierend auf den individuellen Bedürfnissen und Fragen der Nutzer. \\
Sicher ist auch der Gedanke interessant das System als Ansprechpartner im Support-System dienen zu lassen, um häufige Fragen zu beantworten und einfache Probleme zu lösen, bevor sie an menschliche Support-Mitarbeiter weitergeleitet werden.

Durch die Integration eines Large Language Models kann die Plattform nicht nur ihre analytischen Fähigkeiten erweitern, sondern auch die Benutzerunterstützung erheblich verbessern, indem sie kontextbezogene und personalisierte Hilfestellungen bietet. Dies führt zu einer insgesamt besseren Benutzererfahrung und einem effizienteren Betrieb der Plattform.

\section{Fazit}
Insgesamt zeigt diese Ideenskizze, dass die Kombination von Blockchain-Technologie, elektronischen Portfolios und modernen KI-Methoden einen innovativen Ansatz darstellt, der die Effizienz, durch automatisierte Prozesse, und Transparenz des Prüfungssystems deutlich verbessert. Dieser Ansatz bietet sowohl für Studierende als auch für Lehrende und Arbeitgeber klare Vorteile und könnte eine zukunftsweisende Alternative zum traditionellen Leistungsnachweis  darstellen. An verschiedenen Hochschulen wurden bereits vergleichbare Formate evaluiert oder etabliert um das vorhandene Zertifizierungssystem zu ersetzen, jedoch noch nicht als Basis für eine vollständige Ablösung des Leistungsnachweises gesehen, das in dieser Ideenskizze der zentrale Baustein ist. Ziel ist es, den (Lern-)Prozess selbst zu begleiten und weniger die absoluten Ergebnisse festzuhalten.

Die Frage, wie weit die Substitution gehen wird, muss im ersten Schritt noch nicht beantwortet werden. Erstes Ziel muss es sein, das digitale Prüfungssystem aufzubauen und prototypisch zu evaluieren, um die Möglichkeiten zu eröffnen. Ein zentrales Element unserer Arbeit im NahLab! ist es daher, die elektronische Portfolio-Plattform, die es Studierenden ermöglicht, ihre Arbeiten zu protokollieren, zu organisieren und zu präsentieren. Der erste Ansatz ist, dies mit den bestehenden Systemen zu orchestrieren und zu evaluieren. 

Besonderer Augenmerk liegt darauf, wie ein nachhaltiges System über die nächsten 20, 50 oder gar 100 Jahre gelingen kann. Sicher ist, dass neben der nachhaltigen Erfassung (aus Sicht des Studierenden, des Dozenten oder auch des Arbeitgebers) der Leistungen auch der nachhaltige Betrieb sichergestellt werden muss. 
\\\\
Selbstverständlich nutzt diese Plattform aus heutiger Sicht moderne Sicherheitsmaßnahmen wie die Verschlüsselung der Daten bei der Übertragung und Speicherung. Auch wird Wert auf die Einhaltung der Datenschutzbestimmungen gelegt. Dennoch, diese Anforderungen und Rahmenbedingungen unterliegen dem Wandel der Zeit. Neben der rein funktionalen Ebene sind auch die Ebenen der politischen, gesellschaftlichen aber auch klimatischen Rahmenbedingungen zu berücksichtigen, die Einfluss auf den Betrieb dieser Systeme haben und vom NahLab! untersucht werden müssen. Hier kann viel Know-How aufgebaut werden für den zukunftsfähigen Betrieb einer kritischen Komponente und kritischer Infrastruktur.

Ein erster Schritt bei der Konzeption ist die Förderung offener Schnittstellen. Die Implementierung von standardisierten API-Schnittstellen wird eine nahtlose Integration mit externen oder bestehenden Systemen und Lernmanagementplattformen gewährleisten, und die Wahl offener Formate oder Standards wird ebenfalls einen wesentlichen Einfluss auf den Entwicklungsprozess haben. Nicht zu vernachlässigen sind auch die Anforderungen, die durch "public code" und "green code" formuliert werden können. Ansätze wie beispielsweise aus dem Verein XHochschule\footnote{siehe \href{https://xhochschule.de/web/home}{https://xhochschule.de/web/home}} sind hier sicher ein guter Startpunkt.

Auch neue Wege sollen gesucht werden, insbesondere in der Auswertung der Leistungen. Der Einsatz eines Large Language Models (LLM) wie GPT-4 erweitert die analytischen Fähigkeiten der Plattform und bietet kontextbezogene Hilfestellungen. Das LLM kann komplexe Datenanalysen durchführen, Vorhersagen treffen und individuelle Unterstützung bieten, was die Benutzerfreundlichkeit und Effizienz des Systems weiter verbessert. Durch die Automatisierung von Supportanfragen und die Bereitstellung interaktiver Tutorials wird die Lernkurve für die Nutzer deutlich flacher und beherrschbarer. Dennoch müssen auch in diesem Kontext die Persönlichkeitsrechte der Studierende gewahrt werden und eine stetiges Überwachungsszenario ist auszuschließen. 

Die Blockchain-Zertifizierung stellt sicher, dass alle Zertifikate fälschungssicher und öffentlich überprüfbar sind, so dass  das Vertrauen der Arbeitgeber in die vorgelegten Zertifikate gestärkt wird. Dies schließt auch Teilleistungen ein und fördert die berufliche Relevanz der erbrachten Leistungen.
Auch wird eine transparente und faire Bewertung von Leistungen ermöglicht, die nicht für einen akademischen Titel ausreichen. Da diese Leistungen dennoch einen Wert  haben, können sie für zukünftige Arbeitgeber einen Mehrwert darstellen. Daher muss die Plattform auch für nicht rein akademische Abschlüsse geöffnet werden. Eine weitere Öffnung ist auch dahingehend interessant, da die Zertifizierung einen eigenen Dienst darstellt und Firmen oder anderen Institutionen angeboten werden kann, bei Ihnen erbrachten Leistungen über die Blockchain der HAW zu zertifizieren.\footnote{Auch dieser Text wurde begleitend mit Hilfe des Editors ChatGPT und einer Rechtschreibprüfung erstellt.}

\bibliographystyle{IEEEtran}
\bibliography{references}

\end{document}